 \documentclass[final,1p,times,twocolumn]{elsarticle}

\usepackage{graphicx}
\usepackage{amssymb}

\newcommand{\be}{\begin{equation}}
\newcommand{\ee}{\end{equation}}
\newcommand{\ba}{\begin{eqnarray}}
\newcommand{\ea}{\end{eqnarray}}

\journal{Chemical Physics Letters}

\begin{document}

\begin{frontmatter}


\title{Thermodynamically consistent Reference Interaction Site Model
theory of the tangent diatomic fluid}

\author{G.~Muna\`o} 
\author{D.~Costa\corref{cor1}}
\ead{dino.costa@unime.it}
\author{C.~Caccamo}

\cortext[cor1]{Corresponding author}

\address{Dipartimento di Fisica and
CNISM (Consorzio Nazionale Interuniversitario
di Struttura della Materia) \\ Universit\`a
degli Studi di Messina, 
 Contrada Papardo---98166 Messina, Italy}

\begin{abstract}
Thermodynamic and structural properties of the tangent diatomic fluid
are studied in the framework provided by the Reference
Interaction Site Model (RISM) theory, coupled with a Modified Hypernetted Chain
closure. The enforcement of the internal thermodynamic
consistency of the theory is described in detail. The results we obtain
almost quantitatively agree with available or newly
generated simulation data.
We envisage the possibility to extend the consistent 
RISM formalism to generic,
more realistic molecular fluids.
\end{abstract}


\end{frontmatter}



Tangent diatomics, 
constituted by two identical
hard spheres whose centre-to-centre distance 
is equal to the diameter $\sigma$ of each sphere,
provide a relatively simple prototype model for molecular systems. 
They can be considered as a member of two more general classes,
one constituted  by  
hard dumbbells (or fused hard spheres), where
each sphere composing the molecule can have a different $\sigma$ diameter 
and the bond length (elongation) 
is generally a fraction of $\sigma$, and the other one
 constituted by 
two or more freely jointed hard spheres, a model widely used
as a basic representation for chain-like molecules.

The structural and thermodynamic properties of the tangent diatomic fluid 
including its phase behavior, have been widely analyzed until recently
in terms of both computer simulations and liquid state theories.
Early Monte Carlo calculations 
were carried out by 
Freasier {\it et al.}~\cite{Freasier:75,Freasier:76}. 
In Ref.~\cite{Tildesley:80} Tildesley and Streett have used
the Monte Carlo  pressure
to evaluate the constants entering
an empirical analytic expression that accurately fits
the compressibility factor as a function of the density.
More recent simulations can be found in 
Refs.~\cite{Nezbeda:89,Jackson:91,Yethiraj:93,Yethiraj:90}.
The fluid-solid equilibrium 
has been investigated via Monte Carlo simulations and 
free energy calculations in Refs.~\cite{Vega:92,Vega:08}. 

Theoretical studies have 
involved the Reference Interaction Site Model
(RISM) theory of molecular fluids developed by 
Chandler and Andersen~\cite{Chandler:72}. In particular,
in Refs.~\cite{Yethiraj:93,Duda:01}
the tangent diatomic fluid
has been analyzed in the context of 
the Chandler-Silbey-Ladanyi ``diagrammatically proper'' formulation of RISM 
(CSL,~\cite{Chandler:82}).
Calculations based on the contracted formalism 
known as ``polymer-RISM''
are reported in Ref.~\cite{Yethiraj:90}.
%%%%%%%%%%%%%%%%%%%%%%%%%%%%%%%%%%%%%%%%%%%%%%%%%%
Among other theoretical studies
we mention 
 a closed form analytic theory
for the structural functions~\cite{Morriss:83}, 
an Ornstein-Zernike-type integral equation theory with a  
non-spherical bridge function~\cite{Labik:90},
a modification of the Verlet theory for hard spheres~\cite{Labik:91},
a scheme based
on the Born-Green-Yvon integral equation theory~\cite{Taylor:94}.
The equation of state of tangent diatomics has been
 investigated via different 
formalisms (see~\cite{Jackson:91,Boublik:76,Dominik:05}
and references therein).

\bigskip

The aim of this Letter is to present 
the first (to the best of our knowledge) thermodynamically 
consistent study 
of the tangent diatomic fluid in the framework 
provided by the RISM
theory. 
We document  
the accuracy of thermodynamic and structural properties 
obtained through such a scheme and envisage
further applications to more realistic molecular
fluids.

\bigskip

The RISM formalism~\cite{Chandler:72}
 is a matrix generalization of
the Ornstein-Zernike equation 
of simple fluids~\cite{Hansen}, relating
the set of site-site
pair distribution functions
to the corresponding direct correlation functions.
The molecular geometry enters the theory
through a matrix of intramolecular correlations
that takes into account the rigid bonds among the 
various interaction sites of the molecule 
(see 
Ref.~\cite{Monson:90} for a 
review of the method and applications). 
%%%%%%%%%%%%%%
Because of the symmetry of the tangent diatomic molecule
all site-site correlations are equal,
so that the RISM equation assumes the simple form
in $k$ space:
\ba\label{eq:rism}
h(k) = [w(k)+1]^2\,c(k)+ 2\rho [w(k)+1]c(k)h(k)\,.
\ea
In Eq.~(\ref{eq:rism}) $\rho$ is the molecular number density,
$h(r)=g(r)-1$ and  $c(r)$
are respectively the pair 
and direct correlation functions among any of the two sites
of different molecules, and  $g(r)$ is the site-site
radial distribution function.
The  function $w(k)$, that takes 
into account the intramolecular correlations,
is written as:
$w(k)=\sin(kL)/kL$, where $L \equiv \sigma$ is the bond distance
among the two spheres composing the model.

The RISM equation must be complemented
by a closure relation, that is usually taken to be the
 Percus-Yevick (PY) or the 
Hypernetted Chain (HNC)~\cite{Chandler:72,Hirata:81}.
Going beyond these basic approximations,
we suggest here to adopt 
a Modified Hypernetted Chain (MHNC)
closure~\cite{Rosenfeld:79}; specifically,  
we assume that the well known  
exact expression for the radial distribution function 
of an atomic fluid~\cite{Hansen},
%%%%
\be\label{eq:mhnc}
g(r)=\exp[-\beta v(r) +h(r)-c(r)+E(r)]\,,
\ee
%%%%%%%%%%
may be employed 
for the site-site structural functions
of a molecular fluid.
In the equation above
$v(r)$ is the interparticle potential,
$\beta=1/k_{\rm B}T$  (where $k_{\rm B}$ is the Boltzmann constant and
$T$ the temperature) and $E(r)$ is the ``bridge function''. 
The quotation marks are used in this context because
the theoretical framework for 
a rigorous definition of the bridge diagrams
is provided
by the CSL proper formulation of RISM~\cite{Chandler:82}
(see also the recent developments in Refs.~\cite{Duda:01,Lue:95}).
As in the original MHNC scheme for atomic
 fluids~\cite{Rosenfeld:79}, 
we then approximate $E(r)$ by $E_{\rm HS}(r)$, namely  
the bridge function 
of a hard sphere
fluid  of some effective packing 
fraction $\eta_{\rm HS}$, in the
parametrization of simulation data provided by Verlet
and Weis~\cite{VW}.

In the MHNC scheme for atomic fluids,  $\eta_{\rm HS}$ is adjusted
to enforce the thermodynamic consistency of the theory,
as  for instance by requiring
the equality between
the virial and the compressibility equations of state.
For  a generic molecular fluid the virial equation of state
cannot be deduced in terms of site-site radial
distribution functions~\cite{Cummings:81};
 however, for hard dumbbells,
and hence for the model at issue,
an expression for
the excess free energy per particle has been derived 
by Lowden and Chandler~\cite{Lowden:75} in terms
of an integral over $\sigma$ of 
the value of $g(r)$ at contact, $g(\sigma^+)$:
\begin{equation}\label{eq:virialike}
  \frac{\beta A^{\rm ex}}{N}= 
8 \pi \rho \int_{0}^{\sigma}d\sigma'(\sigma')^{2}
g(r=\sigma'^+;\rho,\sigma') \,.  
\end{equation}
This expression for $A^{\rm ex}$ leads, 
upon derivation with respect to the density, 
to a 
``virial-like'' equation of state,
that can be made 
equal (with a suitable choice of $\eta_{\rm HS}$) 
to the pressure calculated from  
the compressibility route:
\begin{equation}\label{eq:compr}
\frac{1}{\beta}\frac{\partial\rho}{\partial P} = 1+ \rho h(k=0) \,.
\end{equation}
%%%%%%%%
The coupled equations~(\ref{eq:rism}) and~(\ref{eq:mhnc})
have been solved through standard numerical
methods. We employ a discrete grid of
8192 points, with a $r$ spacing $\Delta r=0.005\sigma$.
Complementary calculations
with $2^{15}$ points and $\Delta r=0.002 \sigma$
show no appreciable difference with the smaller grid.
In order to enforce the thermodynamic consistency,
the calculations are repeated at each density
for different values of $\eta_{\rm HS}$
till the pressures coming
from Eq.~(\ref{eq:virialike}) and~(\ref{eq:compr})
coincide within a $\sim 2$\% numerical accuracy.

\begin{figure}
\begin{center}
\begin{tabular}{c}
\includegraphics[width=6.0cm,angle=-90]{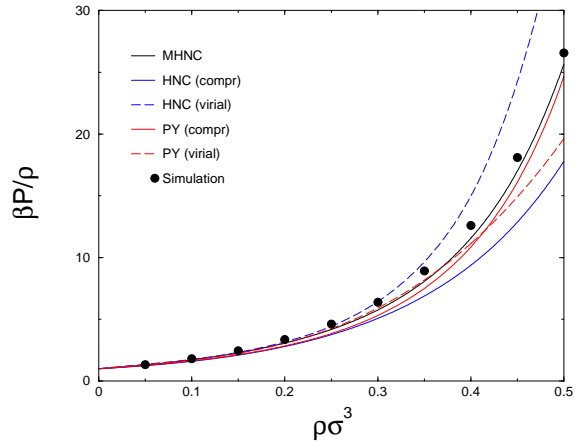} 
\end{tabular}
\caption{MHNC, HNC and PY compressibility factor 
of the tangent diatomic fluid.
Monte Carlo data are taken from Ref.~\cite{Tildesley:80}. 
}\label{fig:1}
\end{center}
\end{figure}  

We report in Fig.~\ref{fig:1}
the RISM/MHNC compressibility factor $\beta P/\rho$
for the tangent diatomic fluid,
gauged against 
Monte Carlo data~\cite{Tildesley:80}, and 
in comparison with the HNC approximation,
(corresponding to the assumption $E(r)=0$ in Eq.~(\ref{eq:mhnc}))
and the PY closure (in which one assumes  $c(r)=0$ outside 
the hard-sphere core).  
As visible, 
the thermodynamically consistent MHNC reproduces
quite well the Monte Carlo equation of state over practically the
whole fluid density range (the freezing threshold of the model
being estimated at $\rho\sigma^3 \sim 0.53$~\cite{Vega:92}).  
The HNC virial-like 
and compressibility equations of state  bracket the simulation data;
as an example, they exhibit
an inconsistency of about 20\% of their average value at $\rho\sigma^3=0.4$.
The HNC pressure obtained from a closed expression 
derived in Ref.~\cite{Singer:85} (not shown in the Figure)
is close to, but systematically
above, the virial-like equation results.
Both PY routes lay systematically
close one to each other and slightly
underestimate the simulation  results.
%%%%%%%%%
In order to complete our picture,
the comparison with other microscopic theories
shows that the MHNC approach and the 
compressibility route in the Born-Green-Yvon (BGY) theory
(see Figure~5 of Ref.~\cite{Taylor:94})
share the same level of accuracy for the pressure
up to $\rho\sigma^3 \sim 0.4$, whereas 
the BGY virial route underestimates systematically
the simulation data from $\rho\sigma^3 \sim 0.3$ onwards. 
The CSL approach
reproduces quite well the Monte Carlo
equation of state 
(see Figure~3 of Ref.~\cite{Yethiraj:93}), although 
it is slightly less predictive than the MHNC in 
the high density fluid regime.

\begin{figure}
\begin{center}
\begin{tabular}{lr}
\includegraphics[width=5.2cm,angle=-90]{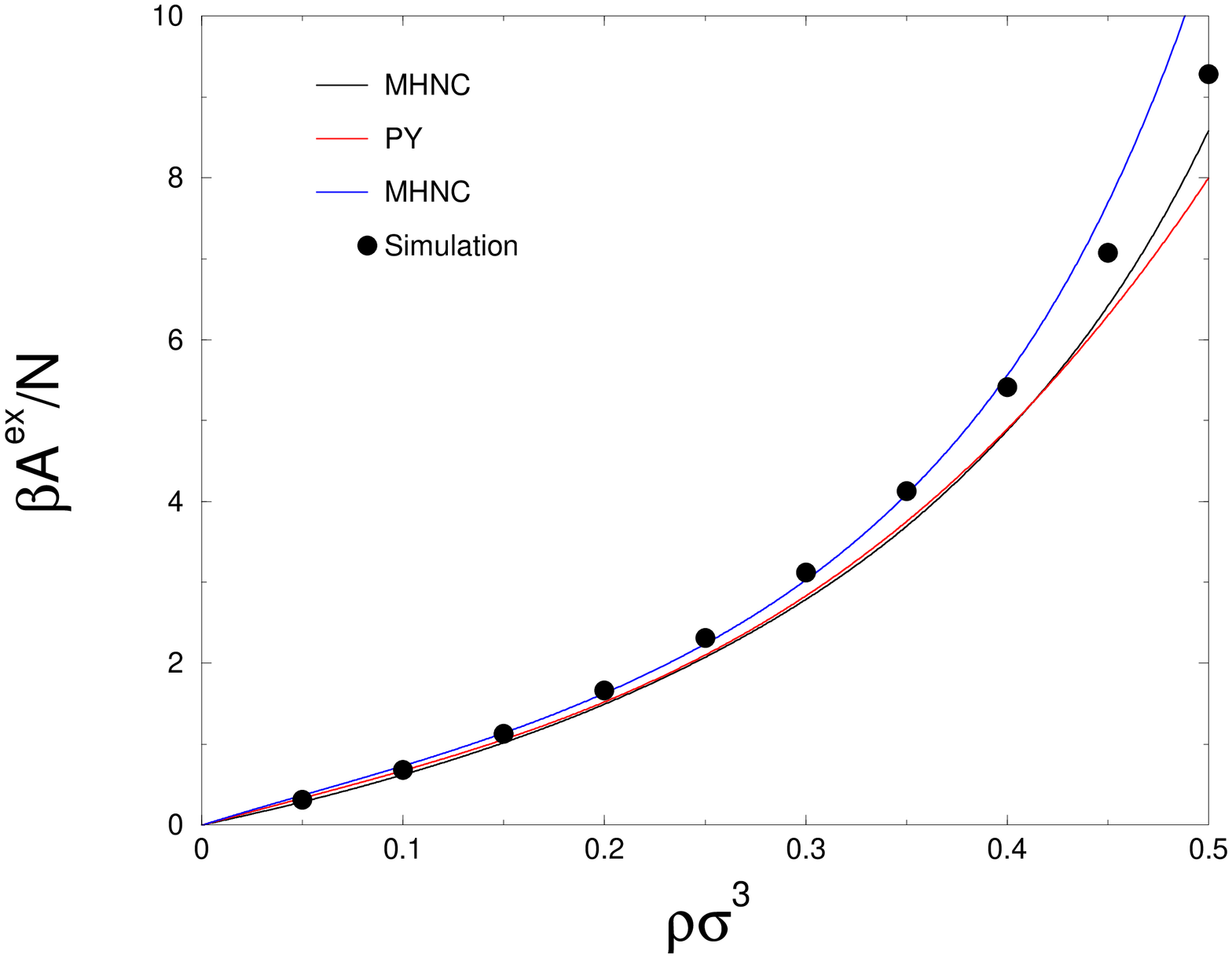}&
\includegraphics[width=5.2cm,angle=-90]{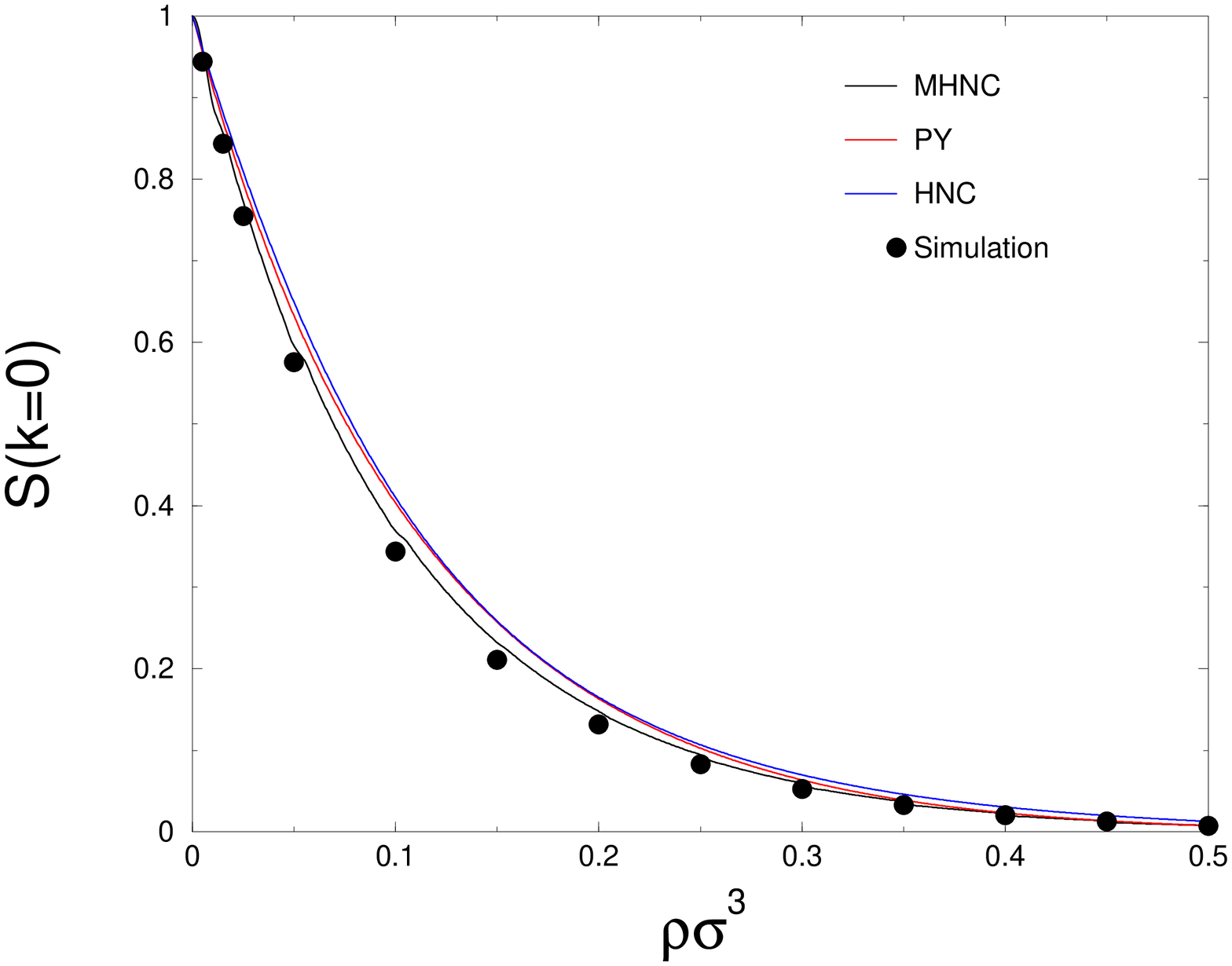}
\end{tabular}
\caption{Theoretical and 
simulation free energy and $k\to 0$ limit
of the structure factor.
MHNC, HNC and PY results (lines) are calculated according
to Eq.~(\ref{eq:virialike}) for the free energy
and Eq.~(\ref{eq:compr}) for $S(0)$.
Simulation data (symbols) are calculated according
to the analytical expressions derived in Ref.~\cite{Tildesley:80}
from an accurate fit of the Monte Carlo pressure.
}\label{fig:2}
\end{center}
\end{figure}

The quality of our results can also be appraised 
from 
Figure~\ref{fig:2}, where we display the MHNC input functions
for the consistency procedure, namely
the excess free energy calculated from Eq.~(\ref{eq:virialike})
and the $k\to 0 $ limit of the static structure factor, $S(0)$
which constitutes the right hand side of Eq.~(\ref{eq:compr}).
The MHNC results are presented along with the HNC and PY predictions,
and compared with the corresponding analytical functions
derived in Ref.~\cite{Tildesley:80} from an accurate fit 
of simulation data for the pressure. We observe that, although all closures
give a relatively accurate reproduction of simulation data, 
the integration of the compressibility and the derivation of the free energy 
eventually give rise to the relative spread of 
theoretical predictions for the pressure
already shown in Fig.~\ref{fig:1}.

\begin{figure}
\begin{center}
\begin{tabular}{lr}
\includegraphics[width=5.2cm,angle=-90]{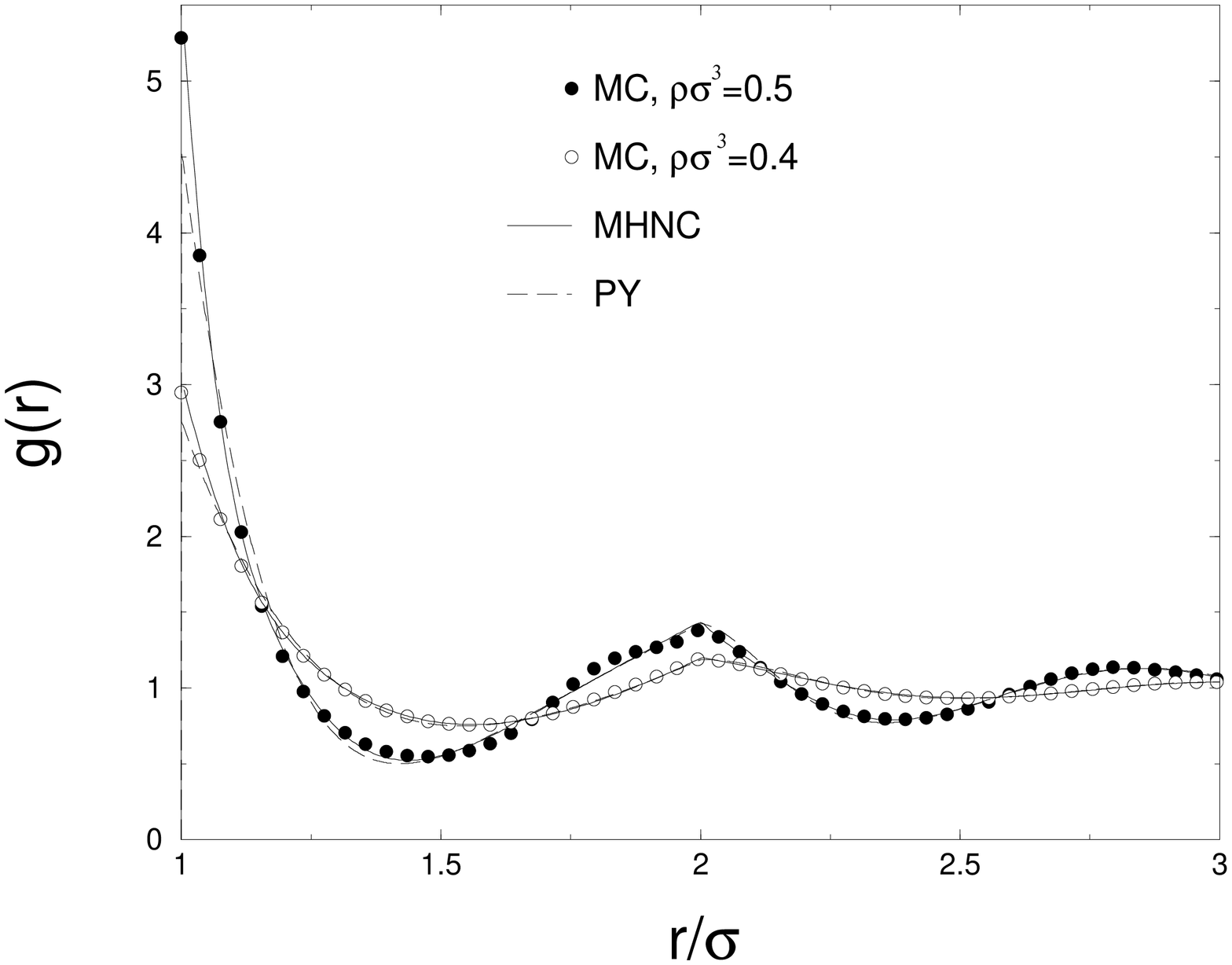} &
\includegraphics[width=5.2cm,angle=-90]{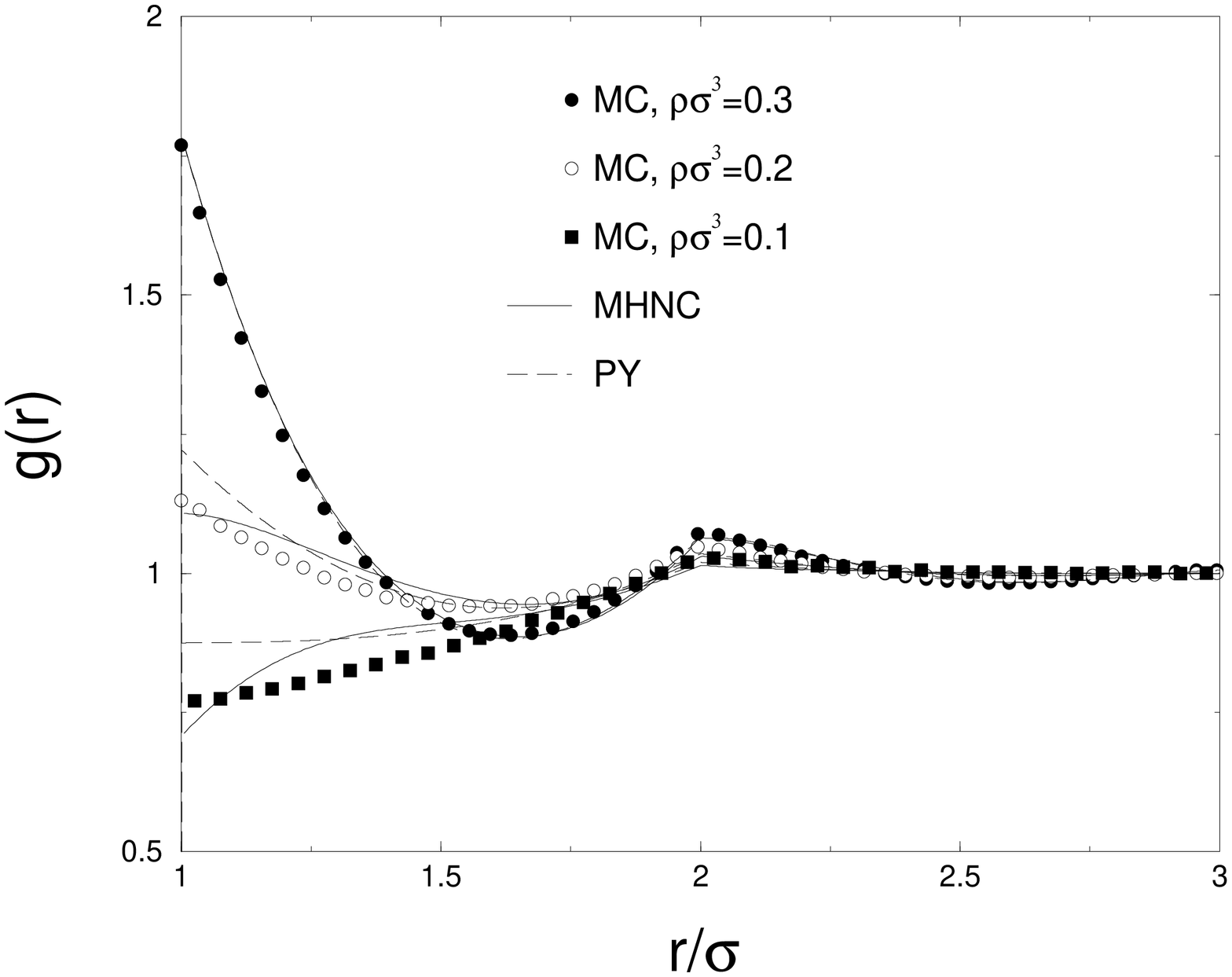} 
\end{tabular}
\caption{MHNC (full lines), PY (dashed lines) 
and Monte Carlo (symbols) site-site
radial distribution functions for the 
tangent diatomic fluid at various densities.  
Theoretical predictions lay close to
the corresponding simulation results 
and are superimposed at $\rho\sigma^3=0.3$.
}\label{fig:3}
\end{center}
\end{figure}

MHNC and PY predictions for
 the site-site radial distribution function $g(r)$
are displayed in Figure~\ref{fig:3}
and compared with Monte Carlo data generated in this work.
Standard Monte Carlo simulations
are generally carried out
on samples composed of $N=500$ molecules
enclosed in a cubic box
with periodic boundary conditions.
We have checked in these conditions
the negligible influence of the box size
and of the spatial mesh of structural functions
through several runs with 4000 particles
and by using spatial grids as fine as $0.001\sigma$ spacing.
It appears from Figure~\ref{fig:3}
that in the intermediate-to-high density regime ($\rho\sigma^3 \ge 0.3$) the 
MHNC correlations are practically superimposed to the 
simulation results.
Only at the highest density investigated, 
$\rho\sigma^3=0.5$, 
the  ``exact'' $g(r)$ displays
a shoulder before the cusp at $r=2\sigma$
preluding to the freezing of the fluid, whereas
the MHNC is less sensitive to this emerging feature.
%%%%
At $\rho\sigma^3=0.1/0.2$, the theoretical correlations 
agree with simulation data for distances around the 
cusp onwards, whereas they appear slightly distorted
in the region $\sigma < r < 2\sigma$. This  artifact is produced by   
too high a value of $\eta_{\rm HS}$, 
necessary to impose the thermodynamic consistency,
As a result, 
the contact value of $g(r)$ is slightly underestimated
in the dilute fluid regime.
The overall quite good reproduction of 
$g(r)$ at contact shown in Fig.~\ref{fig:3}
is noteworthy (the discrepancy with
simulation data barely exceeding $\sim 5$\% in the worst case investigated, 
i.e. at $\rho\sigma^3=0.1$), since $g(\sigma^+)$ constitutes 
an important parameter entering not only Eq.~(\ref{eq:virialike}),
but also other approximate expressions for the equation of state
of tangent diatomics~\cite{Taylor:94}. 
As visible from the figure, the MHNC 
and the PY predictions are indistinguishable
at $\rho\sigma^3=0.3$. The PY is sligthly 
less accurate than the MHNC outside this 
density range, and in particular, 
as already observed in Ref.~\cite{Yethiraj:93},
 tends to overestimate
(underestimate) the contact value of $g(r)$ at lower (higher) densities.
In comparison with our predictions,
results from BGY~\cite{Taylor:94} and CSL~\cite{Yethiraj:93}
theories show a better agreement with simulations data
only in the low density regime.
 
The body of theoretical results  presented in this Letter
appears of remarkable accuracy,
 especially in consideration of the relative
simplicity of the scheme adopted. 
The thermodynamically consistent MHNC
improves on previous RISM results 
and exhibits an accuracy similar to
the more rigorous but comparatively more
complex CSL formalism.
%%%%%%%%%
We have also
several  evidences testifying that the 
MHNC approach positively predicts 
the thermodynamic and structural properties
of hard dumbbells for generic elongations other
than $L=\sigma$~\cite{gmunao:08}.
%%%%%%%%%
The present study paves the way 
for further applications to more realistic systems.
In fact, although  a consistency 
procedure for a generic molecular model cannot hinge on   
Eq.~(\ref{eq:virialike}), 
since this relationship holds strictly for the hard-dumbbell fluid,
other consistency schemes, based e.g. on the ``energy route''
to thermodynamics,  can be implemented
in such cases. Investigations on these topics are
currently underway.

\end{document}